**High sensitivity fluid energy harvester.**

Amit R. Morarka,[a]* Subhash V. Ghaisas[a]
*[a] *Corresponding author: Department of Electronic Science, Savitribai Phule Pune University, Pune, INDIA. Fax: +91-020-25699841; Tel: +91-020-25696060;*
*E-mail: amitm@electronics.unipune.ac.in, amitmorarka@gmail.com*
[a] *Department of Electronic Science, Savitribai Phule Pune University, Pune, INDIA.*
*Fax: +91-020-25699841; Tel: +91-020-25696060; E-mail: svgunipune@gmail.com*

†Appendix available: [Measurement of natural frequency of the cantilever, Fabrication of Copper coil, Fabrication of testing rig, Fabrication and characterization of Droplet dispenser unit.]

**Abstract: -** An ambient energy harvesting device was design and fabricated. It can harness kinetic energy of rain droplets and low velocity wind flows. The energy converted into electrical energy by using a single device. The technique used by the device was based on the principles of electromagnetic induction and cantilever. Readily available materials were characterized and used for the fabrication of cantilever. Under the laboratory conditions, water droplets having diameter 4mm and wind with speed 0.5m/s were used as the two distinct sources. Without making any changes in the geometry or the materials used, the device was able to convert kinetic energy from both the sources to provide voltage in the range of 0.7-1VAC. The work was conceptualized to provide an autonomous device which can harness energy from both the renewable energy sources.

## 1.0 Introduction

Ambient energy harnessing for low power electronic sensors/sensor nodes and instrumentation has attracted many workers over the past two decades. The potential use of the ambient energy has been always concentrated around the low power sensor/sensor nodes which are installed in remote locations for monitoring certain physical parameters. These sensors are installed at the locations where regular human access is not possible but the sensors should keep functioning and transmitting data flawlessly to a remote control station. Though the sensor and its unit are robust, over a period of time the power supply economically becomes unaffordable. To tackle with these problems, energy harvesting through renewable energy source is the first step. It should not only be renewable but it also needs to be abundant in nature. Also the device used to harness it should be easy to fabricate through the existing industrial manufacturing processes.

Various types of energy harvesters have been proposed by many researchers. Their working principles are usually based on electromagnetism, piezoelectricity, thermovoltaics, photovoltaics, biofuel cells, etc. Out of these, piezoelectricity and electromagnetism have gathered attention. Around the globe many workers are in pursuit to create new techniques and fabricate cost efficacious designs to harness low density energy sources which are available in abundance surrounding us. To harness such an abundant low density energy source a system which will be deployed to do this needs to be sensitive towards the changes caused by the low density energy transfer between the source and the system.

Out of the many reported devices for harnessing ambient energy sources, mechanical energy [1, 2] involve while handling devices like cell phones, handy video cameras, etc. was harness to provide power to charge their own batteries. There work was based on harnessing impact energy arising due to shocks occurring while the devices were transported. Energy harnessing device was designed, fabricated and studied using piezoelectric (PZT) material as the transducer and a small steel ball. Due their simple design, the fabrication was also easy and provided efficient way to convert mechanical energy from the bouncing steel ball into electrical energy. They could achieve a maximum efficiency of 35% for converting mechanical energy into electrical for the ball jumping from a height of 5mm. Ambient energy sources available in nature were also explored by many researchers. A single rain droplet could produce [3, 4] approximately 2μJ to 1mJ of energy during normal to downpour of rainfall respectively. The device that they fabricated could convert only the kinetic energy of water droplet into electrical energy by using piezoelectric material namely Polyvinylidenefluoride (PVDF). The team was able to recover approximately 1nJ of electrical energy and 1μW of instantaneous power using raindrops.
Very few workers have mimicked the nature for designing their energy harvesting devices. These natural geometries if replicated into energy harvesting devices, can easily convert energy from unsteady fluid flows into electrical power which is not the case with the standard turbine designs. Out of the many workers who mimicked the natural designs, few [5, 6] developed PZT energy harvesting devices based on the geometry of a tree trunk and grass. The tree trunk design yielded 1-5μW of power under unsteady wind speed of 1.8-4.3m/s. Designs based on PZT were accounted for 10 times the cost of PVDF based designs. In terms of practical usability, the PVDF based energy harvesters could be used even though they produce 1000 times less output power as compared with PZT. PVDF based harvester due to its flexibility, softness and its usability in uncontrolled environments, could be used for long term functionality in comparison with the brittle PZT.
On another front there were steady developments in ambient energy harvesters using electromagnetism. Application such as energy harvesting from the air ducts for powering wireless sensors was developed by [7]. The device had a wing integrated on a cantilever with four Neodymium Iron Boron (NdFeB) magnets, while the coil was placed as a stator. The device worked with an airflow speed of 1.5m/s giving an output power of 20μW. Weak and strong air flows from the environment were harvested by using a windbelt based vibratory linear energy scavenger [8] and a Helmholtz resonator based generator. It could produce output voltage of 81mV peak to peak at 0.53KHz from an air pressure of 50KPa and 4mV peak to peak at 1.4KHz from an air pressure of 0.2KPa respectively. Wind induced vibration of a stay cable [9] to generate power that can be utilized to drive a wireless sensor node. The fabricated device could generate 233.49mW peak to peak and 27.14mW RMS power output at an input acceleration of 74.8mg. The acceleration value corresponds to the ambient wind induced bridge cable vibrations. Humdinger wind belt electromagnetic energy [10-15] harvester, their patented devices could convert the fluid flows into belt oscillations on which magnets are arranged. Stator coils pickup these flux variations and produce electrical energy.
Amongst the non-linear energy harvesters, [16] has developed a novel concept based on potential wells created by the used of quad arrangement of magnets alongwith piezoelectric (PZT) material bi-morph mounted on a cantilever. It is called as Quad stable Energy Harvester (QEH). Repulsive forces between magnets provided the snap through effect to the cantilever causing it to oscillate over larger amplitudes and those giving high output voltage as compare to Bi-stable Energy Harvester (BEH). QEH based device could harness environment energy and produce output voltage of 0.2V (RMS) for the maximum excitation of $0.1g^2Hz^{-1}$ having bandwidth of 5-200Hz. Even though having novel approach, their device could not provide enough voltage which could overcome forward voltage (0.2V) barrier even of a schottky diode. To scavenge wind energy coming from four directions and having response to wide range of wind speeds from 2-17m/s, [17] demonstrated a piezoelectric based energy harvester. Their 'Arc' shaped cantilever showed a peak output power of 1.73mW at maximum wind speed from a specific mode of oscillation of the cantilever. Rest of the modes was producing power but relatively very less as compare to a mode which was providing maximum power. They demonstrated use of the power generated by two serially connected PZT cantilevers to drive commercially available 18 LED bulbs. Irrespective of their four direction response, the device generated less than half the maximum power from the remaining three modes. On the other hand, collecting weak mechanical vibrations and storing them in electrical energy was done by [18]. Their device had a mechanical switch which was actuated by the same energy source whose energy was scavenged. PZT generated potential was stored in an inductor. Back emf generated from the inductor when the switch was opened and closed was used to overcome the forward voltage drop of the rectifying diodes. Once the current crossed the diodes, it was used to charge a capacitor. Their experimental prototype was able to charge a super capacitor upto a voltage of 1.1V in approximately 900sec. Indeed a novel idea but with a major drawback of a mechanical switch. As a thumb rule, the more the moving parts, the bigger the chances of the failure of the device. The repeated use of the metallic contacts eventually oxidizes and erodes away.

In all the previously reported devices, limitations like complex designs, mechanical switching, very low output voltage, non functionality to harness other ambient energy sources, etc motivated us to approach our problem in a little different way. We conceptualized an idea of harnessing ambient kinetic energies of rain droplets and low velocity (less than 1m/s) wind flows through a single device using only one energy conversion principle (electromagnetic transducer) and with a very simple mechanical design. To the best of our knowledge there does not exist, such device which can convert energy from two completely uncorrelated renewable ambient energy sources such as kinetic energy of rain water droplets and ambient winds. The challenge was to develop a device which could harness from both the sources without the need of any alterations in its geometry or the materials used when it is exposed to any of the two ambient energy sources. To our surprise, the solution came through the real time observation of tree/plant leaf which oscillate when the raindrop hits the leaf or when the wind blows by it. Based on this a device was designed and fabricated using a cantilever made from overhead transparency (OHP) sheets on which a copper coil was mounted while a NdFeB magnet was kept below the coil part of the cantilever.

## 2.0 Experimental

As the proposed concept relies on harvesting energy from uncorrelated renewable energy sources, the device should encompass techniques which can convert energies from two sources into electrical usable power without any addition of new components or by making any changes in its geometrical arrangement.
The working principle of the device was based on the concept of electromagnetism and cantilever.

### 2.1. Fabrication and characterization of cantilever

According to [3], for the laboratory testing purposes, we selected droplets having diameter in the range 3-5mm which will have energy distributed in the range of 2μJ – 1mJ. A wind flow over a cross sectional area of $17 \times 10^{-3}$ m$^2$ and speed of 0.5m/s, will contain kinetic energy [19] approximately equal to 1mJ.
The amount of momentum transfer involved with such energy values will also be of smaller magnitude. Hence to efficiently transfer the energy from droplets and wind into oscillations of the cantilever, the cantilever should have inertia comparable to that carried by the falling rain droplet and the targeted wind speed. One such candidate which was qualified as the material for the cantilever was (OHP) sheet. Chemically it is called cellulose acetate [20]. The commercially available OHP sheet is available in the A4 size and having thickness of 100μm.
A cantilever shaped was taken out from the sheet. The physical parameters of the cantilever are listed in Table -1. The natural frequency and thus the Young's modulus of the cantilever were determined by observing the oscillations of the cantilever as described in appendix [A]. Equation (1) was used to calculate the young's modulus [21] of the material of the cantilever from the natural frequency of its oscillations.

$$f = \frac{1.02}{2\pi} \left[\sqrt{(E/\rho)}(t/l^2)\right] \tag{1}$$

Where f - Natural frequency
E - Young's modulus of the cantilever material
$\rho$ - Density of the cantilever material
t - Thickness of the cantilever
$l$ - Length of the cantilever

Table – 1 Physical characterization of the cantilever

| Sr. No. | Description of parameters | Values/name | Units |
|---|---|---|---|
| 1 | Material | Cellulose acetate (OHP) | - |
| 2 | Cantilever dimensions (length*width*thickness) | (70 x 1.076 x 100e-3)e-2 | m$^3$ |
| 3 | Mass | 0.08113e-3 | Kg |
| 4 | Mass Density | 1352 | Kg/m$^3$ |
| 5 | Natural frequency by measurement | 5.0 | Hz |
| 6 | Natural frequency by simulation | 5.4 | Hz |
| 7 | Young's Modulus using equation (1) | 3.559e09 | Pa |

Natural frequency of the cantilever was determined for various lengths from 0.07m to 0.12m. The obtained data was plotted on graph as frequency (f) versus $(1/l^2)$ and a linear fit was obtained from it. Using the slope of the line and the equation (1), young's modulus of the material was calculated.

### 2.2 Fabrication of copper coil

Since the internal resistance offered by an emf source, such as an induction coil, is low as compared to piezo materials, hence the principle of electromagnetic induction was selected to convert the mechanical energy into electrical energy.
The soul of the device is in its coil. Large number of turns were necessary to achieve high magnitude of induce emf in the coil but without increasing the weight of the coil. The trade off was made by using an enameled copper wire having gauge number AWG 53 as described in appendix [B]. The physical properties of the fabricated coil are depicted in Table – 2.

Table – 2 Physical properties of the fabricated coil

| Sr. No. | Description | Values |
|---|---|---|
| 1 | AWG number 53 | 17μm |
| 2 | AWG number 41 | 70μm |
| 3 | Number of turns | 7000 |
| 4 | Weight of the coil with the small self adhesive tape. | 0.27112gm |
| 5 | Resistance of the coil with the AWG 41 no. wire. | 10.32KΩ |

### 2.3 Testing of harvester under the action of falling droplets

A testing rig was fabricated as discussed in appendix [C]. The cantilever alongwith the coil was mounted on the rig for all further testing.
The droplet dispenser unit was fitted over a stand and exactly aligned over the cantilever arrangement at the height of 0.83m. Figure 1 shows the schematic of the arrangement of the dispenser and the cantilever. The output of the coil was logged as discussed in appendix [C].

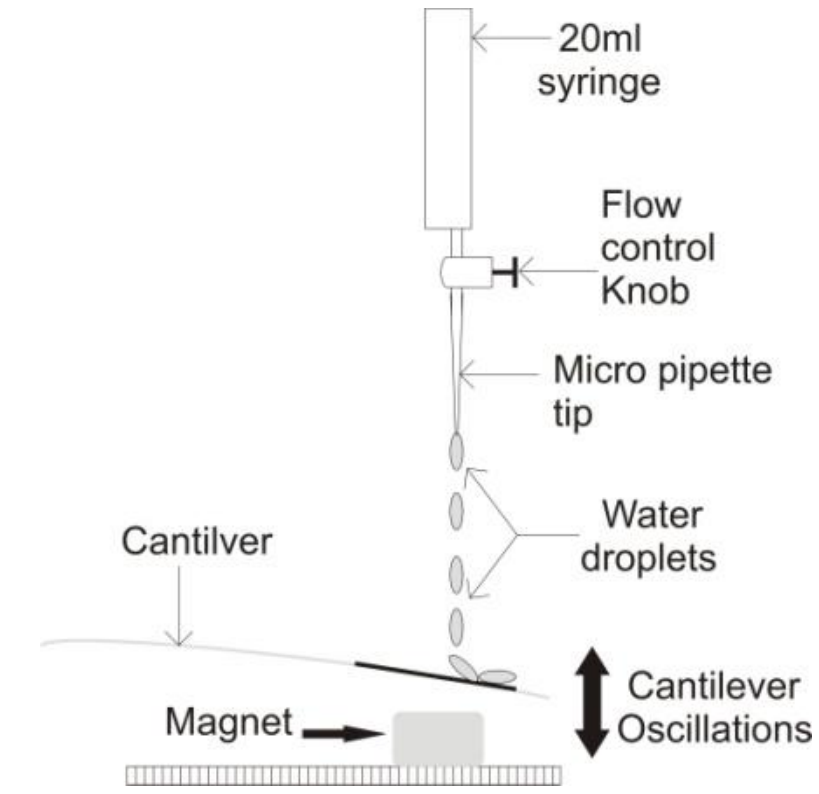

**Figure 1**: Schematic shows the side view of the testing of the harvester arranged under the droplet dispensing unit. The black coloured thin rectangle on the top side of the cantilever is the copper coil. The distance from end of the micropipette tip and the top surface of the cantilever was measured to be 0.83m.

### 2.4 Testing of harvester in ambient wind conditions

The cantilever assembly along with the magnet was used without making any changes to its geometry or in the materials used for its fabrication. The system was exposed to ambient winds generated by two ceiling fans inside the laboratory. The schematic in the Figure 2 shows how the cantilever assembly was kept under the ceiling fans. The distance between the ceiling fans and the surface of the cantilever was 2.66m. The placement of the cantilever system was such that, the wind flow experienced by the system was turbulent in nature from all the directions and having wind speed of 0.5m/s as measured using Lutron's Am-4202 anemometer. The output of the coil was logged as discussed earlier.

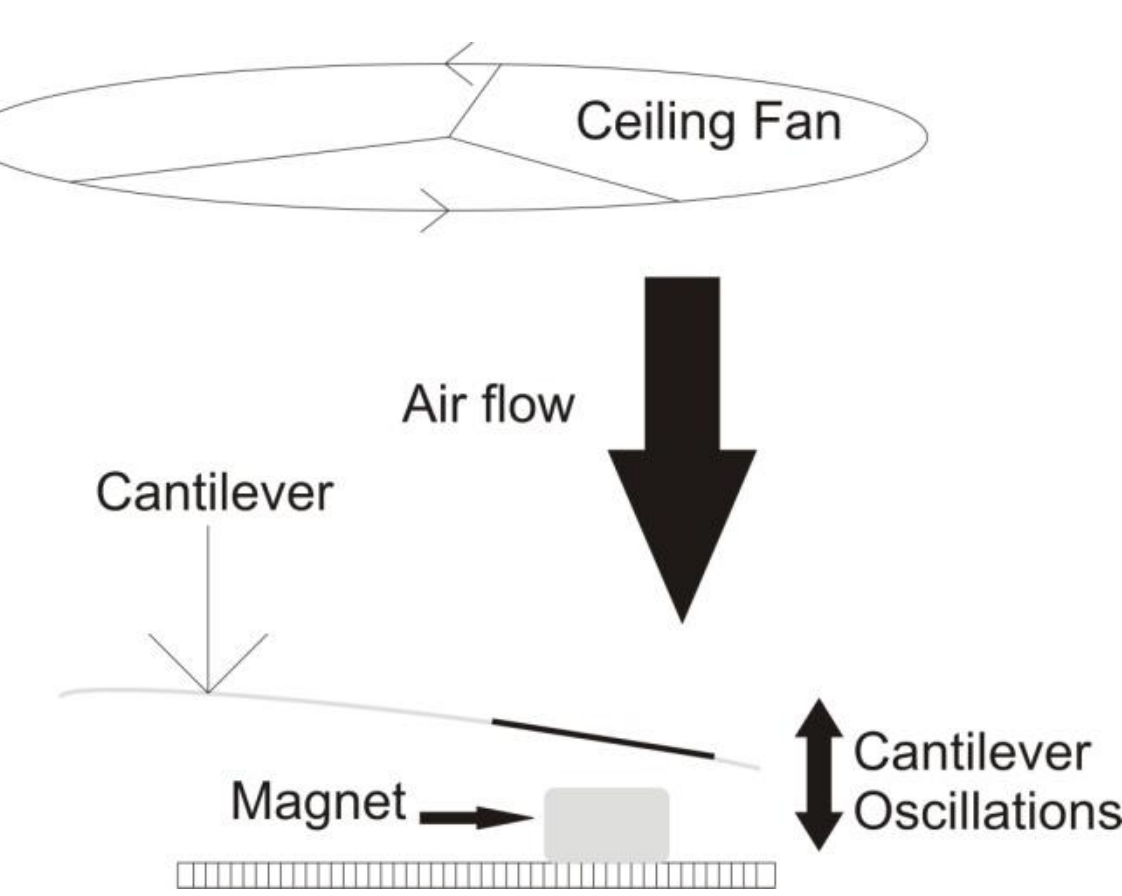

**Figure 2**: Schematic shows the side view of the testing of the harvester arranged under the ceiling fans. Only one fan is shown. Distance between the fan and the surface of the cantilever was measured to be 2.66m.

### 3.0 Results and Discussions

Even though there was a sufficiently good control over the size of the droplet, size characterization of the droplet was needed. The droplet falling out of the dispenser unit was characterized as described in appendix [D].

To determine the Young's modulus of the material of the OHP sheet the cantilever was characterized for its natural frequencies at various lengths. Table – 3 shows natural frequencies of the cantilever for various lengths of the cantilever.

Table – 3 Natural frequencies at various cantilever length without copper coil.

| Sr.No. | Oscillating length (m) | $1/l^2$ (1/m$^2$) | Natural Frequency (Hz) |
|---|---|---|---|
| 1 | 12e-2 | 69.44 | 2.5 |
| 2 | 11e-2 | 82.64 | 2.82 |
| 3 | 10e-2 | 100.00 | 3.33 |
| 4 | 09e-2 | 123.45 | 3.74 |
| 5 | 08e-2 | 156.25 | 5.0 |
| 6 | 07e-2 | 204.08 | 5.98 |

The values of $1/l^2$ and natural frequencies ($f$) were plotted as shown in the Figure 3. Slope of the graph was 0.02651Hz.m$^2$, which was obtained by fitting the curve with linear fit. Equation (1) was rearranged to get,

$$slope = \frac{1.02}{2\pi} \sqrt{\frac{E}{\rho}} \, .t$$

Where $slope = f.l^2$ (Hz.m$^2$) (2)

Thus, the Young's modulus of the material of the OHP transparency sheet was determined from the above equation to be, $E = 3.55e9Pa$

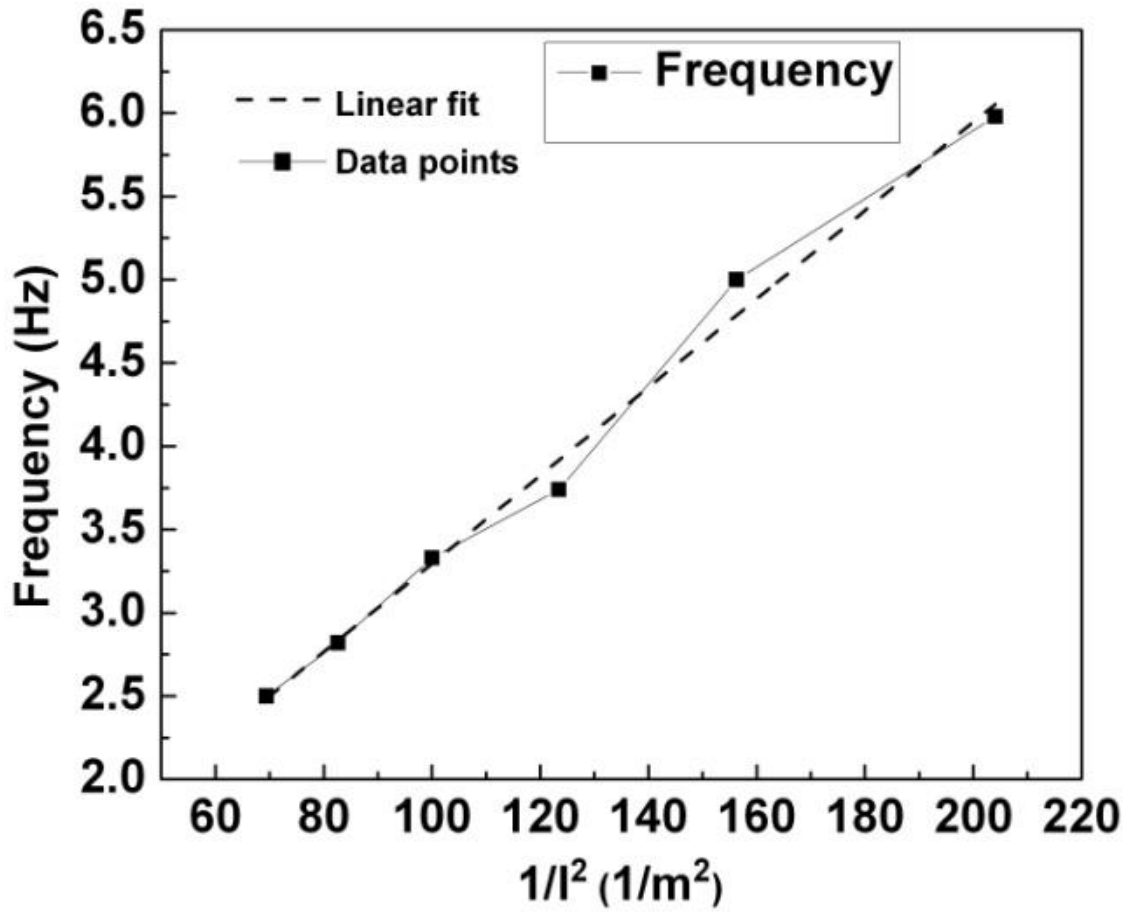

**Figure 3**: Graph of natural frequency versus 1/ (length)$^2$. The linear fit equation was $y = a + bx$. The respective values of slope and the y-intercept was 0.02651 and 0.64376. Pearson's r value for the measured data was 0.99532.

Based on the observations made on the pilot experiments for converting kinetic energy of the falling water droplets, 0.07m length was chosen as the length of the cantilever. The selection of length was based on the fact that the lever arm length should provide enough torque to cause the displacement of the cantilever tip in the range of 4-6mm under the action of the impacting droplet. Thus, the natural frequency of the cantilever for this length was measured to be 5.98Hz. As a copper coil was attached to the free end of the cantilever, its natural frequency was shifted. Equation (3) gives the change in natural frequency [22] of oscillations of the cantilever due to the addition of concentrated mass.

$$\omega = \sqrt{\frac{K}{m}} \quad (3)$$

The new natural frequency of the cantilever with added mass at its free end was measured to be 1.66Hz. The natural frequency of the cantilever with and without added mass was simulated using COMSOL 5.0. Table – 4 shows the results for the natural frequency of cantilever having length of 0.07m with and without added copper coil mass through experiment and simulation.

Table – 4 Natural frequency of the cantilever of length 0.07m with and without added copper coil mass.

| Natural Frequency | Without Copper coil mass | With Copper coil mass |
|---|---|---|
| Simulation COMSOL 5.0 | 5.43Hz | 1.67Hz |
| Experiment | 5.98Hz | 1.66Hz |

Given the limited control over the experimental parameters (thickness uniformity of the cantilever, coil mass distribution and chemical composition of the cantilever), the simulated and experimental natural frequencies are fairly close.
Another speculation can be made for the deviation of the values is that, in the simulations the software algorithm considered the thickness of the cantilever to be perfectly uniform throughout its volume, whereas the OHP transparency sheets are not prepared by any standardized fabrication techniques. Hence, the local density variations in the OHP transparency sheet may yield different observed values of the natural frequency as compared to the simulated values.
The typical value of 1.66Hz as the natural frequency worked out to be practically very useful. The cantilever could execute oscillations as soon as the droplet impacted the tip and returned to its resting position as the driving force resided.
The cantilever-coil assembly along with the magnet was used to test its working to convert the kinetic energy of the falling droplet to electrical energy. Force exerted by a droplet [23] having bigger diameter will also be large. Hence, droplets pertaining to the downpour rainfall conditions were used which were having diameter of 4mm appendix [D]. These were released from the dispenser on to the surface of the cantilever's free end; the impact caused the cantilever to bend down towards the magnet. The coil connected to this end experienced changing magnetic flux through it which in turn induced the voltage in it. The falling, impacting and then breaking of droplets induced oscillations in the cantilever. As long as the cantilever was oscillating, the coil showed induced voltage at the output as shown in the figure 4. Induced AC voltages having peak value of 0.7V and average value of 0.3-0.5V were recorded for each impacting droplet. The obtained voltage can be rectified using schottky diodes (Forward voltage drop = 0.2V). The reported length (0.07m) of the cantilever was responding efficiently to the droplets which were falling at the rate of 1-2 drops per second. As the testing proceeded in time, due to hydrophilic nature of the surfaces of cantilever, coil and the magnet, fine water droplets started to adhere to these surfaces.

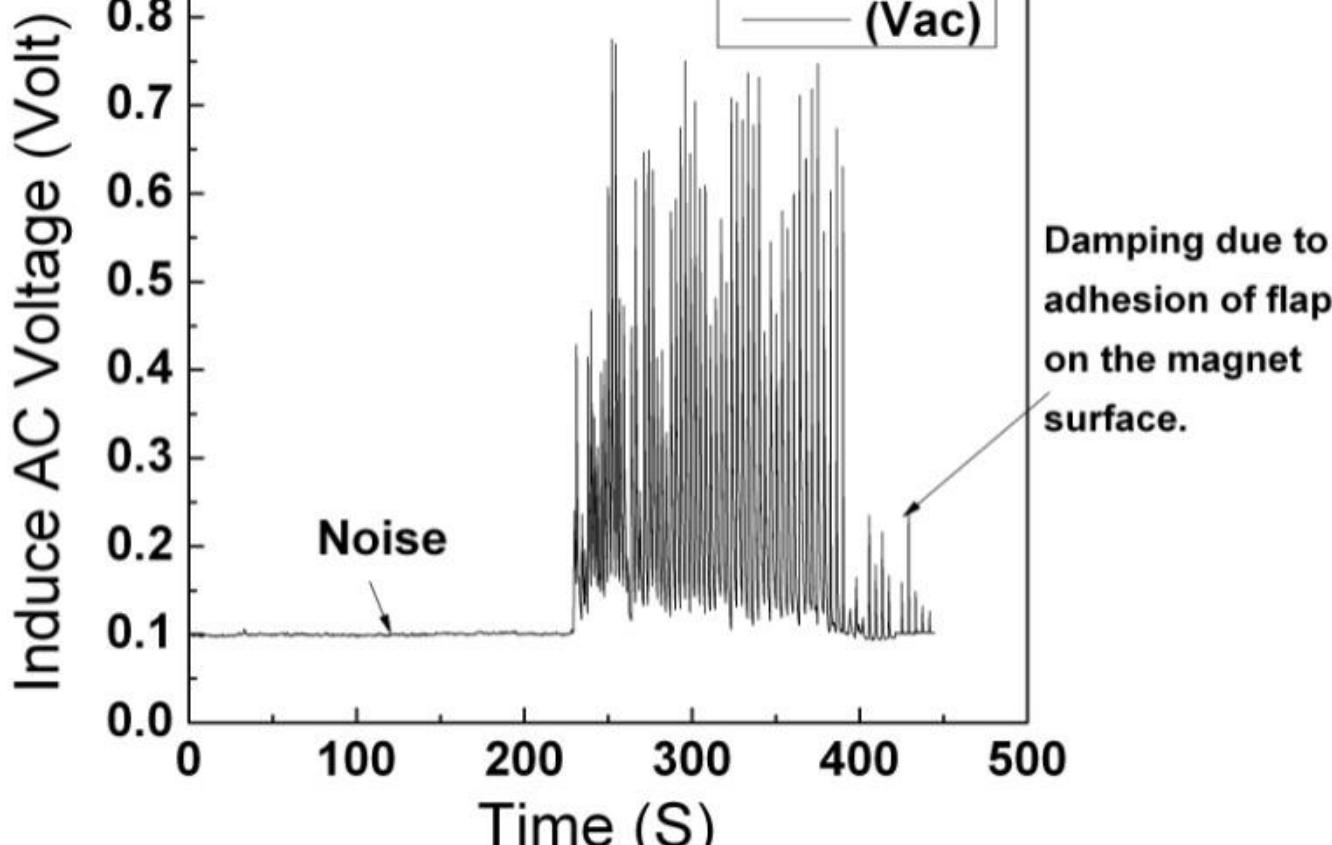

**Figure 4**: Graph of induced AC voltage in the coil with respect to time.

As the experiment proceeded with time, the smaller multiple droplets coalesce on the surface of the cantilever. This caused an increase in the mass at the free end. Due to this the free end which was initially suspended over the magnet got heavier causing it to touch down on the surface of the magnet. With every falling droplet on to the surface of the cantilever, the smaller broken droplets were also coagulating on the surface of the magnet. After certain instant of time, as indicated in the graph of Figure 4, the cantilever surface got adhered on to the surface of the magnet, where the water droplets were acting as the adhesive. There was a small gap left between cantilever and the magnet surface due to their mutual geometrical orientation. Droplets falling on to this were not providing any bigger oscillations to the cantilever but were transferring enough momentum to make it vibrate on its localized position. This can be seen in the graph at the time around 400$^{th}$ second where the magnitude of the voltage was decreased to 0.2V. To solve this problem, a solution was proposed. The surface of the cantilever and the magnet will be coated with superhydrophobic thin film. This way the treated surfaces will inhibit adhesion of any size of water droplets on these surfaces.
As mentioned earlier, the device should be capable of converting from completely uncorrelated renewable energy sources without altering the geometry or the materials used in the device. Hence, the device was taken as it is and exposed to wind flow coming from the ceiling fans inside the laboratory. Due to omnidirectional wind flow having average speed of 0.5m/s in and around the device, the kinetic energy of the wind was transferred to the cantilever, thus causing it to undergo damped non periodic oscillations, which induced AC voltage in the coil. Figure 5 shows the graph of induced voltage recorded at the output of the coil with respect to time.

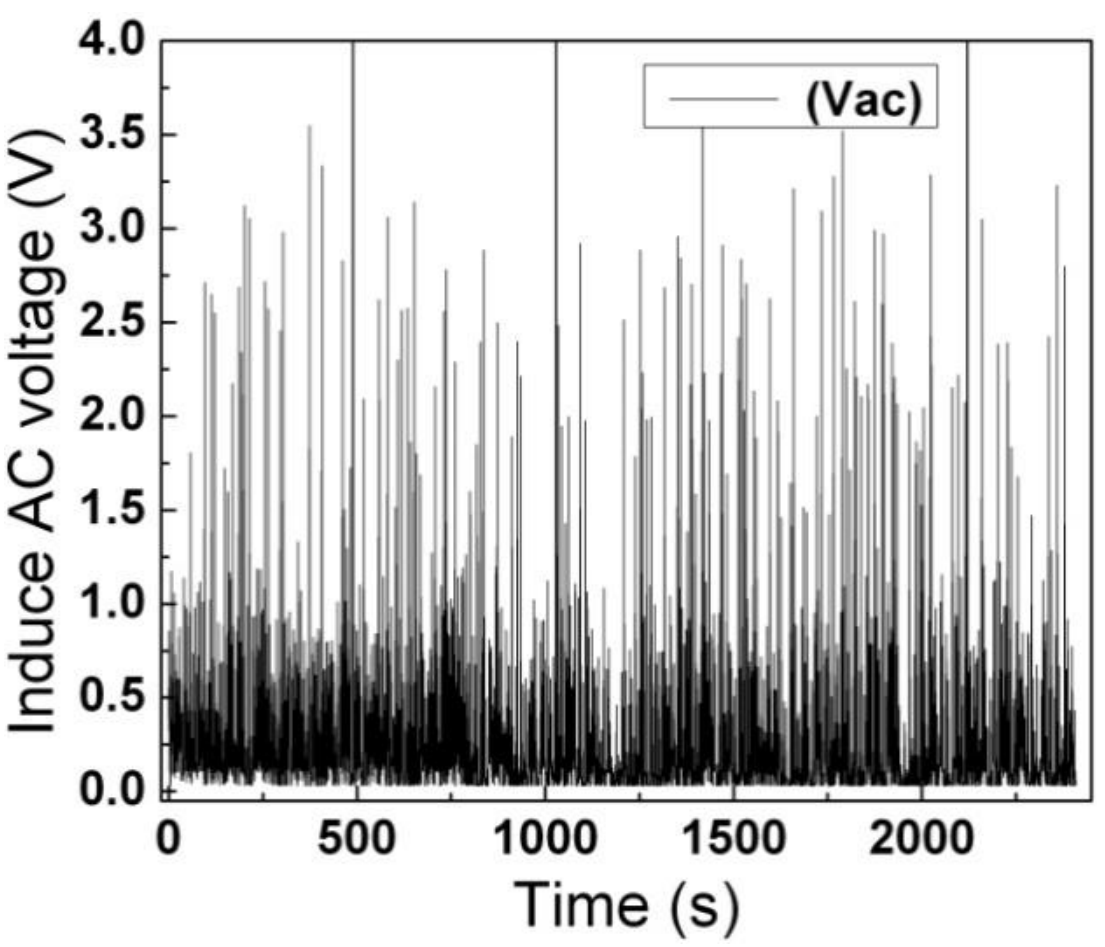

**Figure 5**: Graph of induced AC voltage in the coil with respect to time.

The average magnitude of the induced AC voltage lies in the range of 0.7-1.0V with peak values reaching upto 3V and above. There were very few peaks in the range of 20V due to sudden impulsive movements of the cantilever caused by the changes in the wind directions and magnitude. The oscillations of the cantilever did not have any periodicity as the induced voltage had none. This is due to the fact that even though the ceiling fans generated downward winds, the effective direction acquired by the wind flow was arbitrary as it arrived in the region where the cantilever was placed. This was verified by the use of anemometer. It showed varying wind speeds in all the directions surrounding the cantilever. Hence based on this observation it was concluded that the wind flow was of turbulent type. The device was able to convert some of the turbulent wind kinetic energy into electrical energy almost all the time for which the wind energy was available.

## 4.0 Conclusions

The proposed concept of a device which could truly harness hybrid ambient energy sources was successfully implemented and tested using readily available materials. In-house device was fabricated which could harness energy from two completely independent renewable energy sources. Around 1V AC voltage was obtained autonomously from both the sources of energy without making any changes in the device. The described technique is readily transferable to industrial assembly line for mass production and scalable depending upon the requirement of the energy demand the application. Problems faced while harvesting water droplet kinetic energy were cited and their solutions to these problems were discussed. These solutions had been considered in the on-going work which will be reported in the near future.

**Acknowledgement**

Authors are grateful towards the School of Energy Studies, Physics Department, Savitribai Phule Pune University for providing help in procuring some of the research materials. Authors are highly obliged to Prof. Subramaniam Ananthakrishnan, INSA Senior Scientist, Adjunct Professor, Department of Electronics Science, Savitribai Phule Pune University for providing the laboratory facility and the measuring instruments for this work. Special thanks to Mr. ketan Kanojia and Ms. Aditee Joshi for providing valuable insights while preparing manuscript.

**Funding**

*This work was supported by the Council for Scientific Research and Industrial Research [Grant number: 9/137(545)/2013-EMR-I].*

**Notes and references**

# Appendix

## High sensitivity fluid energy harvester.

### A] Setup for the measurement of natural frequency of the cantilever

As shown in Figure A, the cantilever was fixed from one of its end on to a rigid support. The cantilever was set into oscillations by manually applying a force with arbitrary small amplitude. The oscillations were captured on Sony Cyber shot (model no. DSC-W730) video camera. It had a resolution of 16.1Megapixel with 8X optical zoom. The captured videos were analyzed using a freeware named Avidmux to find out the period of the oscillations of the cantilever.

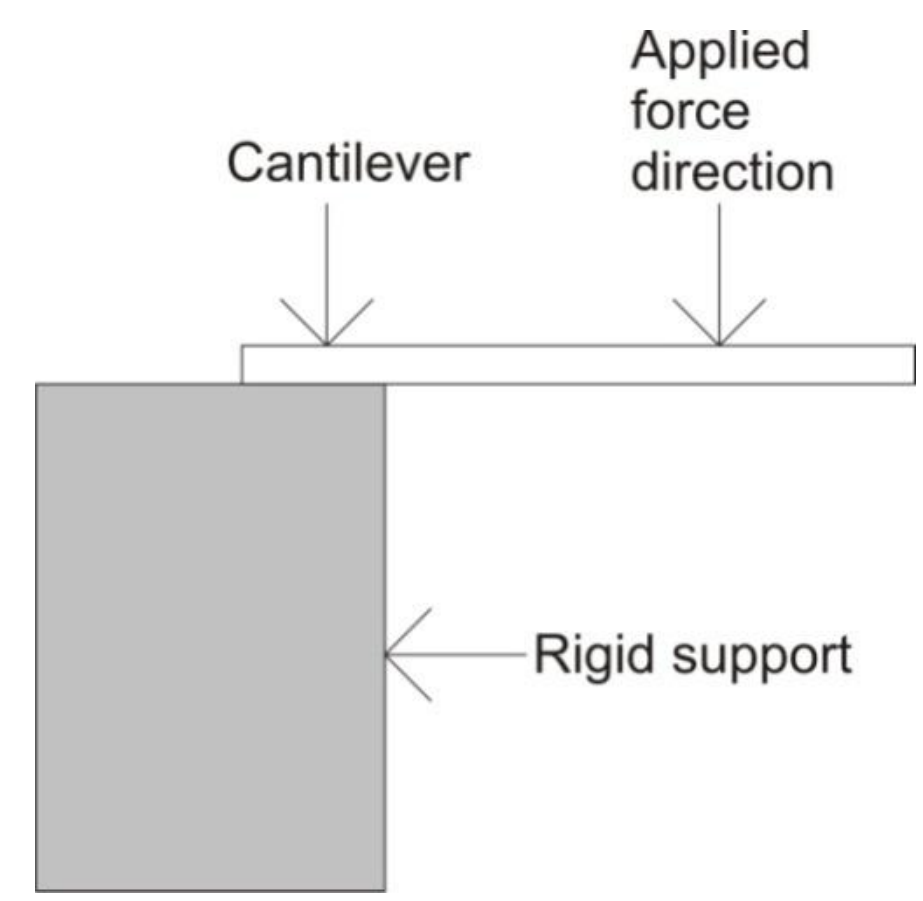

**Figure A**: Schematic shows the side view arrangement of cantilever attached to a rigid support for the measurement of its resonance frequency. The cantilever was attached to the support using self adhesive single sided tape.

### B] Fabrication of Copper coil

Coils were fabricated on an in house developed semi automatic desktop wire winding machine. The coil and the cantilever were held together by a small piece of self adhesive tape. The weight was measured using a digital micro balance from Citizen Scale, model CX165. The two outputs of the coil were connected to AWG 41 number wire using silver paste. This provided good ohmic and robust contacts to integrate the coil with the macro world. Figure B shows the coil assembled on to the cantilever.

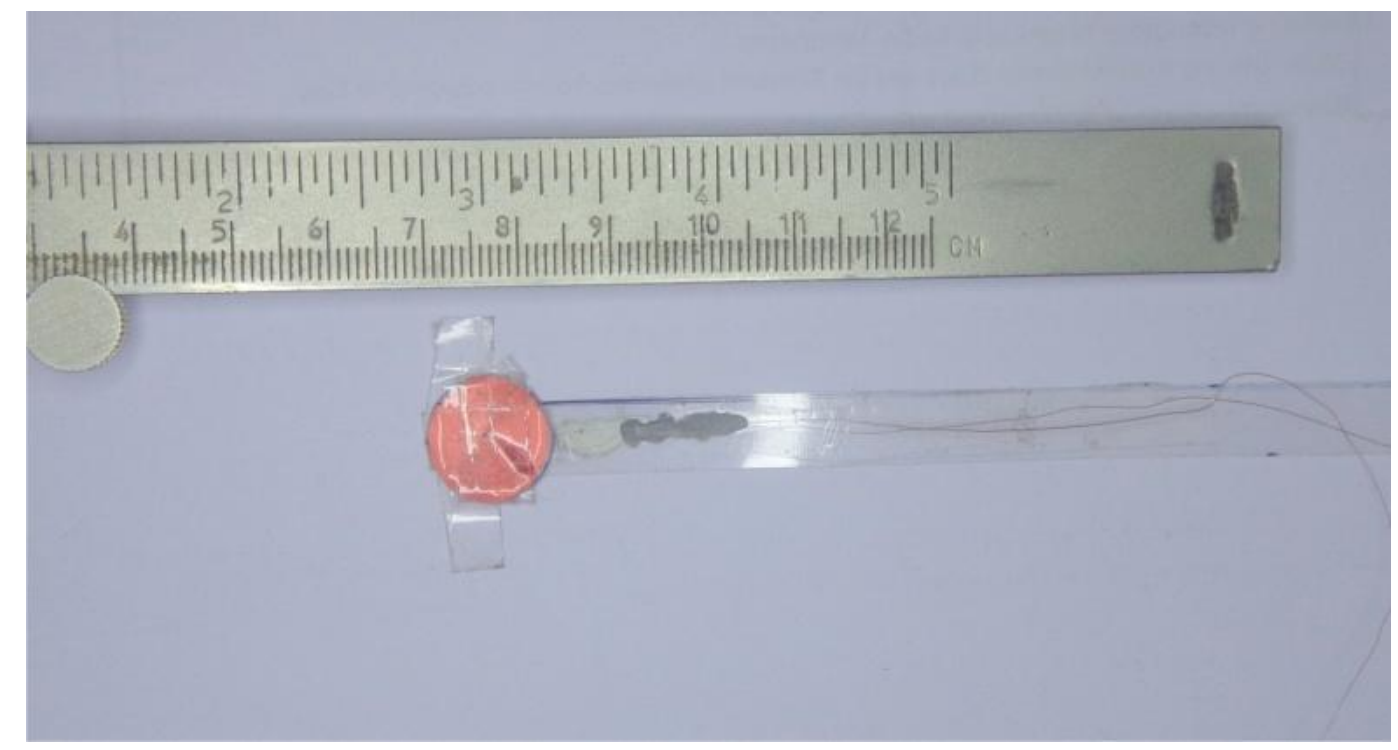

**Figure B**: Image showing the coil assembled on to the cantilever with respect to vernier calliper. Right to the coil is silver paste (arbitrary shape) applied to make ohmic contact between wires having gauge nos. 53 and 41.

## C] Fabrication of testing rig

The rig was made up of acrylic base on which two M3 bolts having 5mm diameter were bolted. On to these bolts, cantilever on one and magnet on to another were affixed. The rig was fabricated in such a way that the positions of cantilever, magnet and the two bolts can be easily reconfigured according to the testing need. Two pole 5mm Pheonix connectors were used to provide connections between coil, capacitors, diodes and the multimeter respectively.
Figure C shows the image of the testing rig. The output voltage from the coil was logged through 6 1/2 Agilent (Model: 34401A) digital multimeter on to PC.

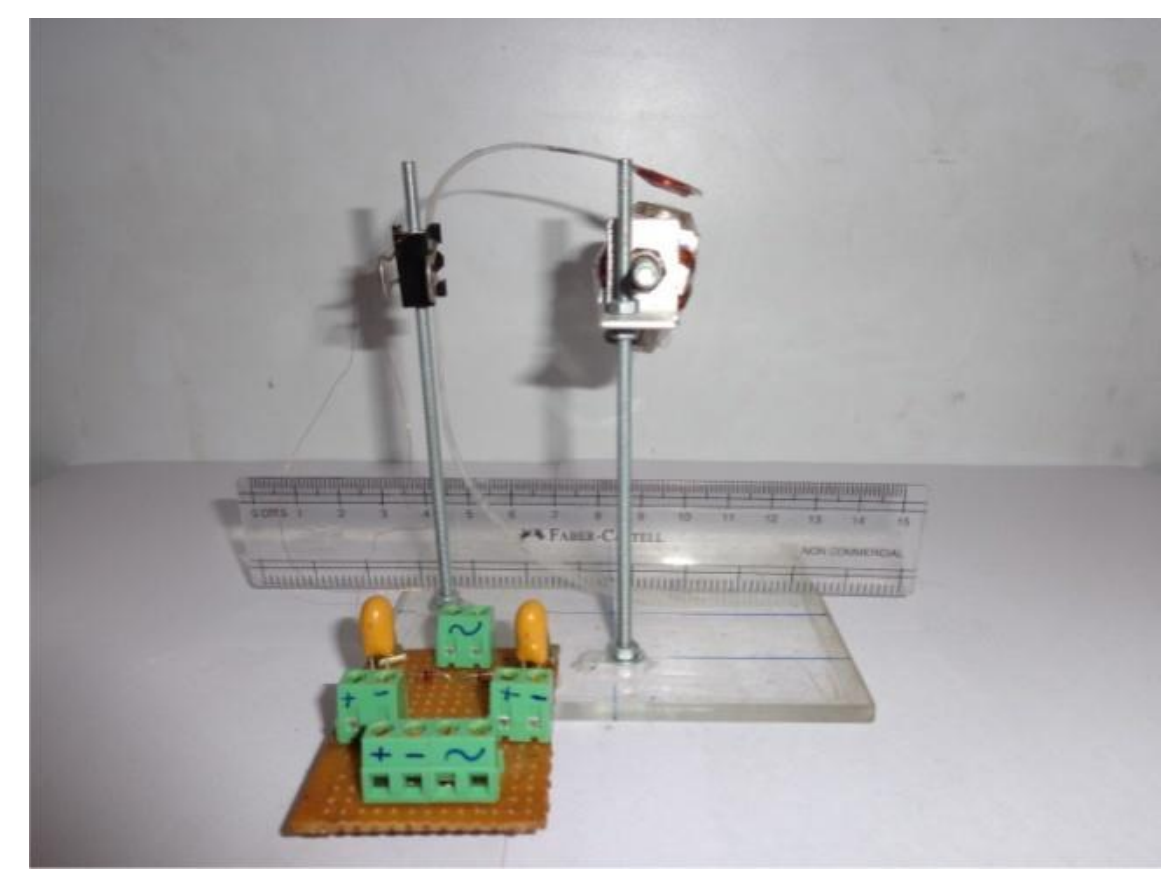

**Figure C**: Image shows the side view of the testing rig. By simply adjusting the nuts on the bolts, the cantilever and magnet can be easily rearranged with respect to each other.

## D] Fabrication and characterization of Droplet dispenser unit

The droplet dispenser is one of the important parts in the experiment. This was the section where the selection of diameter and the rate at which the droplets were released was done. As shown in the Figure D, the droplet dispenser is made up of three different parts. All these three components are simply press fit against each other to make air tight connections. This way any one of these three can be changed to obtain any other desired droplet dispensing rates and or the diameter.

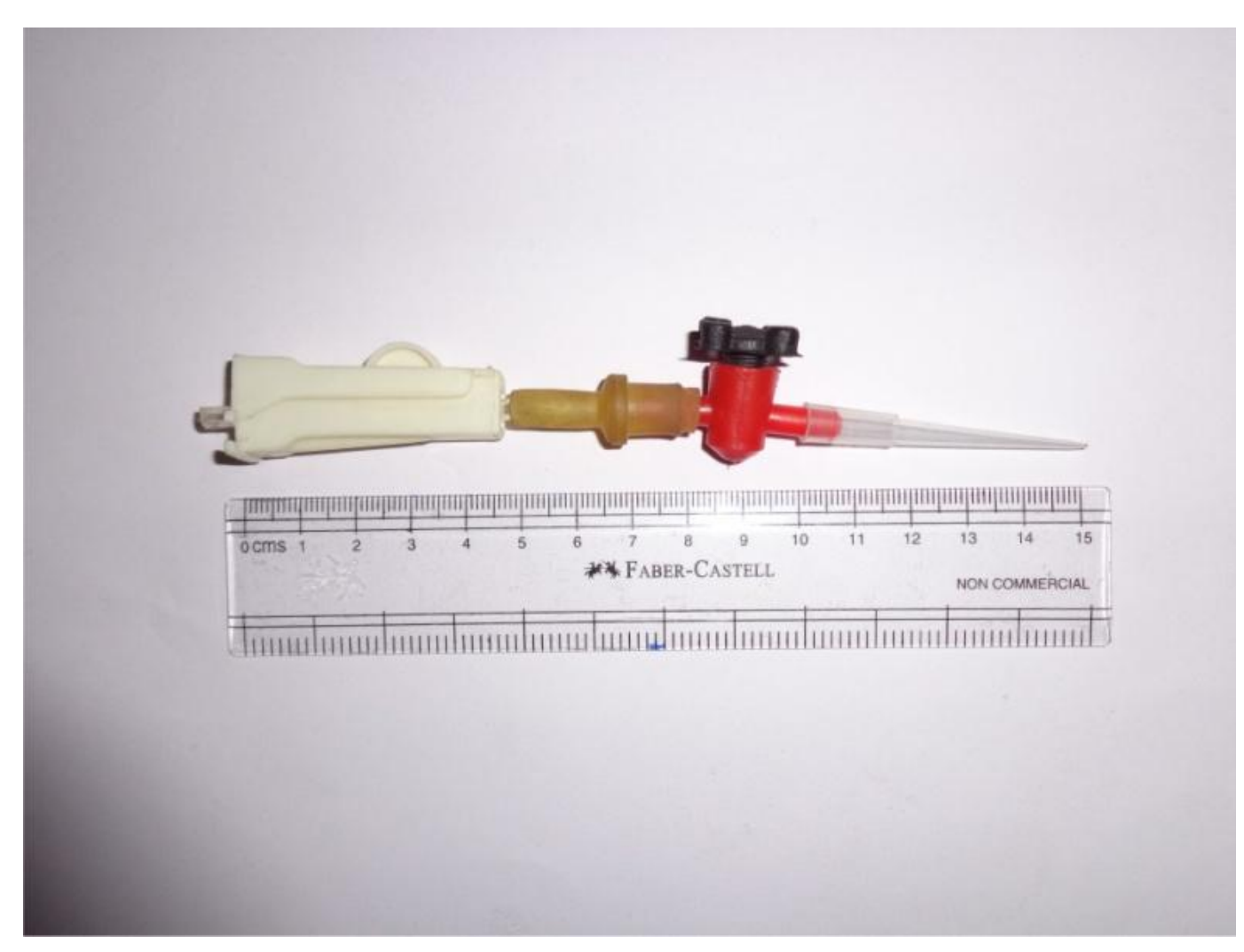

**Figure D**: Image of the droplet dispenser.
From left: IV set flow controller (White), fish tank air flow controller (Red with Black head) and micropipette tip (white Transparent).
These two fluid flow controllers were used in tandem to fine tune the rate at which the individual droplets came out. The last object in the assembly having white colour is the tip of a micropipette. The diameter of the nozzle of the micropipette decides the diameter of the water droplet before it is released.

The dispenser was characterized by depositing a droplet from the dispenser nozzle on to the microbalance under the action of gravity and its mass was measured. Using the equation (I) and from the values of the density of water and its measured mass, the diameter of the droplet was calculated. Table – 1 below shows the values for the characterization of water droplet.

$$V = 4/3(\pi r^3) \qquad \text{(I)}$$

Where $V$ = Volume of the sphere
$r$ = Radius of the sphere

Table – 1 Characterization of water droplet.

| Mass of water droplet (Kg) | Density of Water[1] at 25$^{o}$ C (Kg/m$^3$) | Volume of the water droplet (m$^3$) | Diameter of the water droplet (m) |
|---|---|---|---|
| 0.03269e-3 | 997.0480 | 32.78e-9 | 3.97e-3 |

## Notes and references